\documentstyle[aps,preprint]{revtex}
\newcommand{\sgn}{\mbox{sgn}}
\newcommand{\nonun}{\nonumber}
\newcommand{\djk}{\delta_{jk}}
\newcommand{\bnor}{\left(\frac{\pi}{L}\right)^2}
\newcommand{\bnr}{\frac{\pi}{L}}

\newcommand{\del}[2]{\delta_{\sigma_{#1}\sigma_{#2}}}

\begin{document}


\draft

\title{Multicomponent
Calogero Model of $B_N$-Type
Confined in Harmonic Potential}
\author{Takashi Yamamoto\thanks{
{\it E-mail address}
: yam@yukawa.kyoto-u.ac.jp
}
}
\address{%
Yukawa Institute for Theoretical Physics,
Kyoto University, Kyoto 606, Japan}
\maketitle
\begin{abstract}
A new one-dimensional model,
the
multicomponent
Calogero model
of
$B_N$-type
confined in the harmonic potential,
is introduced.
The Lax pair
of this model is determined, and
then
a set of functionally independent conserved operators
are constructed.
Moreover, the energy spectrums of the above model
are obtained by three different methods.
\end{abstract}

%
%
%
%

There has been a recent burst of interest
in a family of one-dimensional
many body systems
with the
$1/r^2$-type interaction.
For comprehensive reviews of
these
systems see Refs. \cite{rev1,rev2}
and references therein.
The two-body interacting potential of
these systems are
supposed to
be
translationally invariant.
Mathematically speaking,
this property reflects the invariance under the action of
the Weyl group of type $A_{N-1}$,
where $N$ is the number of particles.
In the case of continuous models
without internal degrees of freedom,
Olshanetsky and Perelomov
have studied corresponding
systems associated with other
Weyl groups.
See the survey paper Ref. \cite{O-Pa}.
More recently, it has become apparent
that the models associated with the Weyl group
of type
$B_N$ (or $BC_N$) are important to
describe the one-dimensional physics with boundaries.
For example,
the non-relativistic dynamics of quantum sine-Gordon
solitons in the presence of a boundary
was described by
the Sutherland model of
$BC_N$-type (with $\sinh$-interaction)\cite{K-S94}.
This model is also related to
the physics of the quantum electric
transport in mesoscopic systems\cite{B-R,Caselle}.
The Haldane-Shastry model,
which is the discrete version of
the Calogero-Sutherland model,
with the free boundary
conditions
was defined by generalizing the ordinary
({\it i.e.} $A_{N-1}$-type ) model to
the one associated with the Weyl group of type
$BC_N$ \cite{S-A94,B-P-S}.

It is important to study the application of other
$1/r^2$ models associated with
the Weyl group of type $B(C)_N$
to the one-dimensional physics with boundaries.
For the first step of these directions,
we will introduce and analyze
the multicomponent generalizations of
the Calogero model of $B_N$-type.

Before defining the new model,
we briefly recall the Weyl group of type $B_N$
and its action on the physical space.
In this letter, we
will consider the one-dimensional system
with internal degrees of freedom.
Internal degrees of freedom
means that the particle has
some spin (color) $\sigma\in \Omega$.
Here $\Omega$ is a finite set with
$\#\Omega =r$, that is,
we consider the $r$-component system.
Then $N$-particle system is specified
by
$(q_1 \sigma_1,
q_2\sigma_2,
\cdots,
q_N\sigma_N)$,
where
$(q_1,
q_2,
\cdots,
q_N)\in \mbox{{\bf R}}^N$
are particle coordinates
and
$(\sigma_1,
\sigma_2,
\cdots,
\sigma_N)\in \Omega^N$.
The Weyl group action
can be defined on
both the space of particle coordinates
${\bf R}^N$
and
the space of spins
$\Omega^N$.
Firstly, we define the action on
${\bf R}^N$.
In general,
we consider the group
$W$ of the coordinate transformations
\begin{equation}
  \label{b-transform}
(q_1, q_2, \dots, q_N)
\mapsto
(\epsilon_1 q_{\sigma(1)},
 \epsilon_2 q_{\sigma(2)},
 \dots,
 \epsilon_N q_{\sigma(N)}),
\end{equation}
of
$\mbox{{\bf R}}^N$,
where
$\sigma$
are the elements of the
$N$-th symmetric group
$S_N$ and
$\epsilon_j \in \{\pm 1\}$.
We call $W$ the Weyl group\footnote{%
For the complete definition of
the Weyl groups associated
with the simple Lie algebras, see \cite{humph}.
Among others,
the groups
$A_{N-1}, B_N,$ and $D_N$
(also $C_N$ and $BC_N$)
are of particular interest because
they are related to
quantum '$N$'-body interacting systems.}
of type
\begin{equation}
  \label{weyl-trans}
\begin{array}{ll}
A_{N-1},
\ \ \
&
\mbox{if}\ \
\epsilon_j=1,\ (\mbox{for all}\  j),
\\
B_N, (C_N,\ \mbox{or}\ BC_N,)

\ \ \
&
\mbox{if}\ \
\epsilon_j\in \{\pm 1\},
\\
D_N,
\ \ \
&
\mbox{if}\ \
\epsilon_j\in \{\pm 1\}\ \mbox{and}\
\prod_{k=1}^N \epsilon_k =1.
\end{array}
\end{equation}
Of course, the Weyl group of type $A_{N-1}$ is
same as the symmetric group
$S_N$.

Secondly, we consider the $B_N$-action on $\Omega^N$.
For this purpose, we introduce the operators,
$P_j, P_{jk}$ and $ {\bar P}_{jk}$
which are
defined as follows.
Operators,
$P_j, P_{jk}$ and ${\bar P}_{jk}$,
are
acting only on the spin variable.
The operator
$P_j$
acts on the
$j$-th particle
by
$\sigma_j \mapsto \sigma_j^{\star}\in \Omega$
such that
$P_j^2=1$
({\it i.e.}
$\sigma_j^{\star\star}=\sigma_j$).
The operator
$P_{jk}$
acts by the
permutations
of
$j$-th and
$k$-th spins,
\begin{equation}
  \label{perm-act}
(\cdots, q_j \sigma_j,
\cdots,
q_k\sigma_k,
\cdots)
\mapsto
(\cdots, q_j \sigma_k,
\cdots,
q_k\sigma_j,
\cdots).
\end{equation}
And, the operator ${\bar P}_{jk}$ is defined by
${\bar P}_{jk}=P_jP_kP_{jk}$.
These operators represent
the $B_N$-action on $\Omega^N$.
We note that these operators satisfy
the relations\footnote{%
They also satisfy the
so-called reflection equations\cite{Chere0}
which are the generalized Yang-Baxter equations
associated with the Weyl group of type $B_N$.
},
\begin{eqnarray}
  \label{b-perm}
& &
P_j^2
=
P_{jk}^2
=
{\bar P}_{jk}^2
=
1,\
P_{jk}
=
P_{kj},
\
{\bar P}_{jk}
=
{\bar P}_{kj},
\nonumber
\\
& &
P_j P_k
=
P_k P_j,\
P_j P_{jk}
=
P_{jk} P_k,
\
P_j {\bar P}_{jk}
=
{\bar P}_{jk} P_k
=
P_{jk} P_j,
\\
& &
P_{jk}P_{kl}
=
P_{kl}P_{jl}
=
P_{jl}P_{jk},
\
P_{jk}{\bar P}_{kl}
=
{\bar P}_{jl}P_{jk}
=
{\bar P}_{kl}{\bar P}_{jl},
\nonumber
\end{eqnarray}
where $j, k,$ and $l$ are distinct in the last two formulae.

Let us introduce models
which are multicomponent
generalizations of
the Calogero model of
$B_N$-type.
The Hamiltonians
are
\begin{eqnarray}
{\cal H}_1
&=&
\sum_{j=1}^N p_j^2
+
V_{B_N},
\\
{\cal H}_2
&=&
{\cal H}_1
+
V_{\mbox{{\tiny conf}}},
\end{eqnarray}
where $p_j=-i\frac{\partial}{\partial q_j}$,
and
\begin{eqnarray}
V_{B_N}
&=&
\sum_{j,k=1,\ j\ne k}^N
\left\{
\frac{\lambda(\lambda-P_{jk})}{(q_j-q_k)^2}
+
\frac{\lambda(\lambda-{\bar P}_{jk})}{(q_j+q_k)^2}
\right\}
+
\sum_{j=1}^N \frac{\lambda_1(\lambda_1-P_j)}{q_j^2} ,
\\
V_{\mbox{{\tiny conf}}}
&=&
\omega^2
\sum_{j=1}^N
q_j^2.
  \label{b-hamiltonian}
\end{eqnarray}
Here
$\lambda, \lambda_1 \in \mbox{{\bf R}}$
are the coupling constants and $\omega>0$.
We note that
the $B_N$-action on $\Omega^N$ induces
a $B_N$-action on the operators
$P_j, P_{jk}$ and ${\bar P}_{jk}$.
Therefore,
Hamiltonians ${\cal H}_1$ and ${\cal H}_2$
are $B_N$-invariant, in spite that they
are not invariant under the $B_N$-action
on ${\bf R}^N$.
Then we refer to the model with the Hamiltonian
${\cal H}_1$ (${\cal H}_2$)
as the (confined)
spin Calogero model of $B_N$-type
($B_N$-(C)SC model).
These models are
natural generalizations of the models which
have been studied
in Refs. \cite{M-P,H-W93,U-W-b,V-O-K,V-O-K2,B-V,D-K-V}.
We note that the $B_N$-SC model has been analyzed
by Cherednik\cite{Chere2},
but
his treatment is quite different from
ours and
the Lax pair of this model
has not been given.
The $B_N$-CSC model is new.

Here we give some comments on the above Hamiltonians.
The term
$\lambda(\lambda-{\bar P}_{jk})/(q_j+q_k)^2$
in $V_{B_N}$
expresses the two-body interaction
between the $j$-th particle
and
the ``mirror-image''(we place a mirror at
the origin $q=0$)
of the $k$-th particle.
The term
$\lambda_1(\lambda_1-P_j)/q_j^2$
in $V_{B_N}$
represents the potential of the {\it impurity}
sitting on the origin.
These terms
violate the translational invariance.
Note that
$V_{\mbox{{\tiny conf}}}$
also violates the
translational invariance.

In what follows,
we only treat the Hamiltonian
${\cal H}_2$,
because almost all calculations
necessary for  the Hamiltonian
${\cal H}_1$ are done in similar manner,
or are included in those for Hamiltonian
${\cal H}_2$.

Now,
following Olshanetsky-Perelomov\cite{O-Pa}
and Ujino-Wadati \cite{U-W-b},
we construct a set of
conserved operators for
the $B_N$-CSC model.
To start with, we take the
$2N\times 2N$ matrices
${\bf L}, {\bf M}$
and ${\bf Q}$ as follows;
\begin{eqnarray}
{\bf L}=
\left(
\begin{array}{cc}
L
&
S
\\
-S
&
-L
  \end{array}
\right),
{\bf M}=
\left(
  \begin{array}{cc}
M
&
T
\\
T
&
M
  \end{array}
\right),
{\bf Q}=
\left(
  \begin{array}{cc}
Q
&
0
\\
0
&
-Q
  \end{array}
\right),
  \label{lax-matrix}
\end{eqnarray}
where
$L, M, S, T,$ and $Q$ are
$N\times N$ matrices depending on
$p_j, q_j, P_j, P_{jk}$ and ${\bar P}_{jk}$;
\begin{eqnarray}
L_{jk}
&=&
\djk
\left(p_j\right)
+(1-\djk)
\left(i\lambda \frac{P_{jk}}{q_j-q_k}\right),
\\
M_{jk}
&=&
\djk
\left(
2\lambda\sum_{l=1,l\ne j}^N
\left\{
\frac{P_{jl}}{(q_j-q_l)^2}+\frac{{\bar P}_{jl}}{(q_j+q_l)^2}
\right\}
+\lambda_1\frac{P_j}{q_j^2}
\right)
\nonun \\
& &
\qquad\qquad\quad
+(1-\djk)
\left(-2\lambda \frac{P_{jk}}{(q_j-q_k)^2}\right),
\label{lax-b}
\\
S_{jk}
&=&
\djk
\left(
i\lambda_1\frac{P_j}{q_j}
\right)
+(1-\djk)
\left(i\lambda \frac{{\bar P}_{jk}}{q_j+q_k}\right),
\\
T_{jk}
&=&
\djk
\left(
-\lambda_1
\frac{P_j}{q_j^2}
\right)
+(1-\djk)
\left(-2\lambda \frac{{\bar P}_{jk}}{(q_j+q_k)^2}\right),
\label{lax-t}
\\
Q_{jk}
&=&
\djk
(iq_j).
  \label{lax-pair}
\end{eqnarray}
These operators satisfy the so-called
sum-to-zero conditions\cite{U-H-W}
and the extended Lax equations which
will lead to conserved operators
for the $B_N$-CSC model.
The sum-to-zero conditions
are the following conditions on
${\bf M}$,
\begin{equation}
  \begin{array}{c}
\displaystyle{
\sum_{\mu=1}^{2N}{\bf M}_{\mu\nu}
=
\sum_{j=1}^N M_{jk}+\sum_{j=1}^N T_{jk}
=0},
\ \mbox{for all $\nu$},
\\
\displaystyle{
\sum_{\mu=1}^{2N}{\bf M}_{\nu\mu}
=
\sum_{j=1}^N M_{kj}+\sum_{j=1}^N T_{kj}
=0},
\ \mbox{for all $\nu$},
 \end{array}
  \label{sum-to-zero}
\end{equation}
where
$k=\nu\ \mbox{for}\ 1\leq \nu\leq N,
k=\nu-N\ \mbox{for}\ N+1\leq \nu\leq 2N$.
These are obvious from the formulae (\ref{lax-b})
and (\ref{lax-t}).
The extended Lax equations
for the $B_N$-CSC model are
\begin{equation}
[{\cal H}_2, {\bf L^{\pm}}_{\mu\nu}]
=
\sum_{\sigma=1}^{2N}
(
{\bf L^{\pm}}_{\mu\sigma}{\bf M}_{\sigma\nu}
-
{\bf M}_{\mu\sigma}{\bf L^{\pm}}_{\sigma\nu}
)
\pm
2\omega {\bf L^{\pm}}_{\mu\nu},
\
(\mu, \nu=1, 2, \cdots, 2N),
  \label{lax-eq}
\end{equation}
where ${\bf L}^{\pm}={\bf L}\pm 2\omega{\bf Q}$.
These equations are equivalent to
the following equations
\begin{eqnarray}
\left[{\cal H}_2, L_{jk}^{\pm}\right]
&=&\sum_{m=1}^N(L_{jm}^{\pm}M_{mk}-M_{jm}L_{mk}^{\pm})
+\sum_{m=1}^N(S_{jm}T_{mk}+T_{jm}S_{mk})
\pm
2\omega L_{jk}^{\pm},
\\
\left[{\cal H}_2, S_{jk}\right]
&=&\sum_{m=1}^N(S_{jm}M_{mk}-M_{jm}S_{mk})
+\sum_{m=1}^N(L_{jm}^{\pm}T_{mk}+T_{jm}L_{mk}^{\pm})
\pm
2\omega S_{jk},
  \label{lax-eq2}
\end{eqnarray}
where,
$j,k =1, 2, \cdots, N$, and $L^{\pm}=L\pm 2\omega Q$.
By using the relations (\ref{b-perm}),
we can prove (\ref{lax-eq}).

Let us consider
the
operators
\begin{equation}
  \label{poly-LL}
\mbox{{\bf L}}^{(m_1, n_1, m_2, n_2, \cdots)}
=
(\mbox{{\bf L}}^+)^{m_1}
(\mbox{{\bf L}}^-)^{n_1}
(\mbox{{\bf L}}^+)^{m_2}
(\mbox{{\bf L}}^-)^{n_2}
\dots,
\end{equation}
where
$m_j, n_j\in \{0, 1, 2, 3, \cdots\}$.From
the extended Lax eq. (\ref{lax-eq}),
we can find that
\begin{equation}
  \label{comm-LL}
\left[
{\cal H},
\left(
\mbox{{\bf L}}^{(m_1, n_1, m_2, n_2, \cdots)}
\right)_{\mu,\nu}
\right]
=
\left[
\mbox{{\bf L}}^{(m_1, n_1, m_2, n_2, \cdots)},
{\cal M}
\right]_{\mu,\nu}
\end{equation}
if
$\sum_{j}n_j=\sum_k m_k$.
Then, due to the sum-to-zero condition (\ref{sum-to-zero}),
we see that the operators
\begin{equation}
{\cal I}_n=\sum_{\mu,\nu=1}^{2N}
{\cal O}
\left({\bf L^+}^n{\bf L^-}^n\right)_{\mu\nu},
\quad n=1,\cdots ,N,
  \label{commu-op}
\end{equation}
commute with
${\cal H}$.
Here
${\cal O}$ denotes the Weyl ordered product,
\begin{eqnarray}
  \label{weyl-order}
{\cal O}(X^n Y^n)
&=&
\frac{n!n!}{(2n)!}
\left(
X^n Y^n
\right.
\nonumber
\\
&+&
X^{n-1}Y^nX
+
X^{n-1}Y^{n-1}XY
+
X^{n-1}Y^{n-2}XY^2
+
\cdots
+
X^{n-1}YXY^{n-1}
\nonumber
\\
&+&
X^{n-2}Y^nX^2
+
X^{n-2}Y^{n-1}X^2Y
+
X^{n-2}Y^{n-1}XYX
+
\cdots
+
X^{n-2}YXYXY^{n-2}
\nonumber
\\
&+&
\cdots
\nonumber
\\
&+&
\left.
Y^nX^n
\right).
\end{eqnarray}
Then ${\cal I}_n$'s
are conserved operators.
We shall
give another expression for the conserved operators.
By using the operator
${\cal A}^{\epsilon\epsilon'}
=
(L-S+\epsilon\omega Q)
(L+S+\epsilon'\omega Q),
$
where
$\epsilon, \epsilon'\in \{\pm \}$,
we can rewrite (\ref{commu-op}) in the form,
\begin{equation}
  \label{inv-def}
{\cal I}_n=
\frac{n!n!}{(2n)!}
\sum_{\sum_{j=1}^n(\epsilon_j+\epsilon'_j)=0}
\sum_{j,k=1}^N
\left(
{\cal A}^{\epsilon_1\epsilon'_1}
\cdots
{\cal A}^{\epsilon_n\epsilon'_n}
\right)_{jk}.
\end{equation}
For example,
$
{\cal I}_1
=
\frac{1}{2}
\sum_{\epsilon+\epsilon'=0}
\sum_{j,k=1}^N
({\cal A}^{\epsilon\epsilon'})_{jk}
=
\sum_{j, k=1}^N
\left(
L^2-S^2+\left[L, S\right]-\omega^2 Q^2
\right)_{jk}
=
{\cal H}_2.
$
In the case of the $A_{N-1}$-type
model,
the corresponding
${\cal I}_1$ represents the total momentum.
On the other hand, in this case, the total momentum is not a
conserved operator
because the translational invariance is absent
in the Hamiltonian ${\cal H}_2$. From
the explicit form of ${\cal A}^{\epsilon\epsilon'}$,
we can find that ${\cal I}_n$
is divided into two parts,
\begin{equation}
  \label{form-of-inv}
{\cal I}_n(\{q_j\},\{p_j\})
=
{\cal I}_n^{\mbox{{\tiny poly}}}
(\{q_j\})
+
{\cal I}_n^{\mbox{{\tiny homo}}}
(\{q_j\},\{p_j\}),
\end{equation}
where
$
{\cal I}_n^{\mbox{{\tiny poly}}}
(\{q_j\})
$
is the polynomial of $q_j$'s with the
highest degree term
$\omega^{2n}\sum_j q_j^{2n}$, and
${\cal I}_n^{\mbox{{\tiny homo}}}$
has the form,
\begin{equation}
  \label{homo}
{\cal I}_n^{\mbox{{\tiny homo}}}
(\{q_j\},\{p_j\})
=
\sum_{j=1}^N
p_j^{2n}
+
{\mbox{($q_j$'s dependent part)}},
\end{equation}
which satisfies the homoginuity condition
$
{\cal I}_n^{\mbox{{\tiny homo}}}
(\{\alpha q_j\},\{\alpha^{-1}p_j\})
=
\alpha^{-2n}
{\cal I}_n^{\mbox{{\tiny homo}}}
(\{q_j\},\{p_j\})
$.
Then, due to the formula (\ref{homo}),
we can conclude that the conserved operators
${\cal I}_1, \cdots,{\cal I}_N$
are functionally independent.
In this way we obtained the conserved operators of
the
$B_N$-CSC model.

For later use, we define the operators,
\begin{eqnarray}
  \label{bilinear}
a_j
&=&
\sum_{\mu=1}^{2N}
{\bf L}_{j\mu}^-
=
\sum_{k=1}^N
(L_{jk}^- + S_{jk})
\\
&=&
p_j
-
i\omega q_j
+
i\lambda
\sum_{l=1,l\ne j}^N
\left(
\frac{P_{jl}}{q_j - q_l}
+
\frac{{\bar P}_{jl}}{q_j + q_l}
\right)
+
i\lambda_1
\frac{P_j}{q_j},
\\
a_j^\dagger
&=&
\sum_{\mu=1}^{2N}
{\bf L}_{\mu j}^+
=
\sum_{k=1}^N
(L_{kj}^+ - S_{kj})
\\
&=&
p_j
+
i\omega q_j
-
i\lambda
\sum_{l=1,l\ne j}^N
\left(
\frac{P_{jl}}{q_j - q_l}
+
\frac{{\bar P}_{jl}}{q_j + q_l}
\right)
-
i\lambda_1
\frac{P_j}{q_j},
\end{eqnarray}
where
$j=1, 2, \dots, N$.
These satisfy the following
commutation relations,
\begin{eqnarray}
\left[
a_j, a_k
\right]
&=&
\left[
a^{\dagger}_j, a^{\dagger}_k
\right]
=
0,
\\
\left[
a_j, a^{\dagger}_k
\right]
&=&
\delta_{jk}
(2\omega+M_{jj}-T_{jj}).
\label{non-v-com}
\end{eqnarray}
Then it is easy to see that
the Hamiltonian ${\cal H}_2$ is factorized as
\begin{equation}
\label{fact}
{\cal H}_2
=
\sum_{j=1}^N
a^{\dagger}_j a_j
+
{\cal E},
\end{equation}
where
$
{\cal E}
=
\omega
(
N
+
2\lambda
\sum_{j<k}
(P_{jk}+{\bar P}_{jk})
+
2\lambda_1
\sum_{j} P_j).
$
The operator ${\cal E}$ is related
to the second Casimir operator
of the Lie algebra ${\em so}_{2N+1}$.

Next we will consider the ground state of
the $B_N$-CSC model.
We shall consider two cases which are
characterized by the value of
$\lambda$ and $\lambda_1$.
First, we consider the case
$\lambda, \lambda_1>0$.
In this case, we can easily find the ground state
because in an appropriate representation of spins
the operator
$M_{jj}-T_{jj}$
in the r.h.s. of
eq. (\ref{non-v-com})
is always positive.
Then,
the function
$\Phi^{(0)}(\{q_k\}, \{\sigma_k\})$
which
is a solution to
the equations
\begin{equation}
  \label{g-s-eq}
a_j
\Phi^{(0)}(\{q_k\}, \{\sigma_k\})
=0,
\
(j=1, \cdots, N),
\end{equation}
is the ground state of the $B_N$-CSC model,
since the term
$\sum_j a_j^{\dagger}a_j$ in r.h.s. of the formula
(\ref{fact})
is semi-positive definite.
Note that
eqs. (\ref{g-s-eq}) are equivalent to the
so-called
$B_N$-type Knizhnik-Zamolodchikov equation\cite{Chere1},
\begin{equation}
  \label{b-KZ}
\left(
\frac{\partial}{\partial q_j}
-
\lambda \sum_{l=1,l\ne j}^N
\left(
\frac{P_{jl}}{q_j-q_l}
+
\frac{{\bar P}_{jl}}{q_j+q_l}
\right)
-
\lambda_1
\frac{P_j}{q_j}
\right)
\Phi^{\mbox{{\tiny K-Z}}}(\{q_k\}, \{\sigma_k\})
=
0,
\end{equation}
where we put
$
\Phi^{(0)}(\{q_k\}, \{\sigma_k\})
=
\Phi^{\mbox{{\tiny K-Z}}}(\{q_k\}, \{\sigma_k\})
\prod_{l=1}^N e^{-\frac{1}{2}\omega q_l^2}
$.
We can
construct
$\Phi^{(0)}(\{q_k\}, \{\sigma_k\})$
in the form,
\begin{equation}
  \label{g-s-func1}
\Phi^{(0)}(\{q_m\}, \{\sigma_m\})
=
\phi_{\lambda,\lambda_1}^{(0)}(\{q_m\})
\chi
(\{\sigma_m\}),
\end{equation}
where
\begin{equation}
  \label{one-comp-ground}
\phi_{\lambda,\lambda_1}^{(0)}(\{q_m\})
=
\prod_{1\leq j < k \leq N}
|q_j-q_k|^\lambda
|q_j+q_k|^\lambda
\prod_{l=1}^N
|q_l|^{\lambda_1}
\prod_{n=1}^N
e^{-\frac{1}{2}\omega q_n^2}
\end{equation}
is the Jastrow-type
ground state of the corresponding one-component model,
and
$\chi
(\{\sigma_m\})$
is some function which is invariant under the action of
$P_j$ and $P_{jk}$. Note that the above ground state
is $B_N$-invariant.

Next we consider the case
$\lambda<0$
or $\lambda_1<0$.
In this case,
the ground state
does not need to annihilate by
$a_j$'s.
However, following \cite{M-P,V-O-K2},
we conjecture that the (spin-singlet) state
\begin{eqnarray}
  \label{g-s-func2}
{\tilde \Phi}^{(0)}(\{q_m\}, \{\sigma_m\})
&&
=
\phi_{-\lambda,-\lambda_1}^{(0)}(\{q_m\})
\nonumber
\\
\times
&&
\prod_{1\leq j < k \leq N}
(q_j-q_k)^{\del{j}{k}}
(q_j+q_k)^{\del{j}{k}}
\exp
\left\{
i\frac{1}{2}
\mbox{sgn}(\sigma_j-\sigma_k)
\right\}
\prod_{l=1}^N
q_l
\end{eqnarray}
is the ground state of the $B_N$-CSC model with
$P_j=1$ and $\lambda, \lambda_1<0$.
Note that in the condition
$P_j=1$ the Hamiltonian ${\cal H}_2$ possesses
the $SU(r)$-symmetry.
At least, we can show that
the state
${\tilde \Phi}^{(0)}(\{q_m\}, \{\sigma_m\})$ is
the {\it eigen} state of the Hamiltonian
${\cal H}_2$ with the eigen value,
\begin{equation}
\label{nega-gra-ena}
{\tilde E}_N^{(0)}
=
\omega
\left\{
N
-
2\lambda
N(N-1)
-
2\lambda_1
N
+
2\sum_{\alpha=1}^r
N_{\alpha}^2
\right\},
\end{equation}
where
$N_\alpha$ is the number of particles
with spin $\alpha$ ($\sum_{\alpha=1}^r N_{\alpha}=N$).

We now turn to the spectrum of the $B_N$-CSC model.
We will derive the spectrum by
using three different methods.
The first is the explicit construction
of the excited
states (we call the direct method),
more precisely
the triangulation of the
Hamiltonian ${\cal H}_2$.
The second is the renormalized-harmonic
oscillator (RHO) method\cite{RHO} which is
a variant of
the asymptotic Bethe-ansatz method.
The third is the operator method.
In what follows,
we fix the spin configuration
$\{N_\alpha\}$.

To begin with, we consider
the direct method.
We introduce the
$B_N$-invariant free boson bases
(Notice that the model is confined in the harmonic potential),
\begin{equation}
  \label{excite-cinf}
\psi(\{n_m\})
=
\sum_{\epsilon_1,\cdots,\epsilon_N\in\{\pm1\}}
\sum_{\sigma\in S_n}
\prod_{a=1}^N
H_{n_{\sigma(a)}}(\sqrt{\omega}\epsilon_a q_a).
\end{equation}
Here
$H_n$ is the Hermite polynomial,
and $n_1, n_2, \cdots, n_N$ are
even\footnote{%
Due to the formula
$H_n(-x)=(-1)^n H_n(x)$
for the Hermite polynomial,
summands in the r.h.s.
of (\ref{excite-cinf})
with odd $n_j$
are vanished.
This reflects the $B_N$-invariance.
}
non-negative integers such that
$n_1\geq n_2 \geq \cdots\geq n_N$.
We will fix the ordering $\succ$
of the above basis.
For the two sets of even non-negative
integers,
$n_1\geq n_2 \geq \cdots\geq n_N$
and
$m_1\geq m_2 \geq \cdots\geq m_N$,
we write
$\psi(\{n_j\})\succ \psi(\{m_j\})$
if the first nonvanishing
difference
$n_k-m_k$ is positive.
Then we can show that
with respect to the above ordering
the Hamiltonian ${\cal H}_2$
is triangular
in the basis
$\{{\hat\Psi}(\{n_j\})
=
{\hat \Phi}^{(0)}(\{q_m\},\{\sigma_m\})
\psi(\{n_j\})\}$
where
${\hat \Phi}^{(0)}(\{q_m\},\{\sigma_m\})$
is the Jastrow-type ground state with the ground
state energy ${\hat E}_N^{(0)}$.
Therefore,
the energy spectrum
is obtained by reading
the diagonal
elements labeled in terms
of the quantum numbers
$\{n_j\}$.
The result is
\begin{equation}
  \label{one-energy}
{\hat E}_N
=
{\hat E}_N^{(0)}
+
2\omega\sum_{j=1}^Nn_j.
\end{equation}
We note that $\sum_{j=1}^Nn_j$ is
always the even non-negative integer.
This is the essential difference from
the $A_{N-1}$-case and corresponds to
the existence of {\it mirror} particles.

Next,
we construct
the RHO
solution
which is due to Kawakami\cite{RHO}.
The essence of
the RHO method is that all the interaction effects are
incorporated in terms of the renormalized quantum numbers of
oscillators.
This method is supported by the following observation. From
the formula (\ref{form-of-inv}),
in the asymptotic region
$0\ll q_1\ll q_2\ll\cdots\ll q_N$,
the operators
$\{{\cal I}_n\}$ have the form
$\{\sum_j (p_j^{2n}+\omega^{2n}q_j^{2n})\}$
which are the  conserved operators of the $N$-independent
harmonic oscillators.

We shall consider the case
which corresponds to the
ground state ${\tilde \Phi}^{(0)}$
given by the formula (\ref{g-s-func2}).
In the RHO method,
the energy spectrum ${\tilde E}_N$ is given by
\begin{eqnarray}
  \label{rho-energy}
{\tilde E}_N
&=&
2\omega\sum_{j=1}^N
(m^{(1)}_j+\frac{1}{2}).
\end{eqnarray}
The renormalized quantum number $m^{(1)}_j$
together with $m^{(\alpha)}_j (\alpha=2,\cdots,r)$
are to be determined by the nested equations
(cf. Ref. \cite{Nao94}),
\begin{eqnarray}
  \label{aba-eq}
& &
m_j^{(1)}
=
I_j^{(1)}
-
\sum_{k=1}^{M_2}
\left\{
\sgn (m_k^{(2)}-m_j^{(1)})
+
\sgn (m_k^{(2)}+m_j^{(1)})
\right\}
\nonumber
\\
& &
+
(-\lambda+1)
\sum_{l=1,l\ne j}^{M_1}
\left\{
\sgn (m_j^{(1)}-m_l^{(1)})
+
\sgn (m_j^{(1)}+m_l^{(1)})
\right\}
+
(-\lambda_1+1)
\sgn (m_j^{(1)}),
\\
& &
2\sum_{l=1,l\ne k}^{M_\alpha}
\left\{
\sgn (m_k^{(\alpha)}-m_l^{(\alpha)})
+
\sgn (m_k^{(\alpha)}+m_l^{(\alpha)})
\right\}
+
2\sgn (m_k^{(\alpha)})
+
I_k^{(\alpha)}
\nonumber
\\
&=&
\sum_{j=1}^{M_{\alpha -1}}
\left\{
\sgn (m_k^{(\alpha)}-m_j^{(\alpha-1)})
+
\sgn (m_k^{(\alpha)}+m_j^{(\alpha-1)})
\right\}
\nonumber
\\
&+&
\sum_{n=1}^{M_{\alpha +1}}
\left\{
\sgn (m_k^{(\alpha)}-m_n^{(\alpha+1)})
+
\sgn (m_k^{(\alpha)}+m_n^{(\alpha+1)})
\right\},\ \alpha=2,\cdots,r.
\end{eqnarray}
Here
$I^{(\alpha)}_j \in \{0,2,4,\cdots\}$
is the bare quantum number, and
the
sign function defined by
$\sgn (x)=1 \ \mbox{for} \ x > 0 \ \mbox{and}
 =0 \ \mbox{otherwise}$.
In the above equation,
the quantity
$M_{\alpha}=\sum_{\beta=\alpha}^r N_{\beta}$
was introduced ($M_1=N,\  M_{r+1}=0$).
By substituting the
nested equations (\ref{aba-eq})
to the formula (\ref{rho-energy}),
we obtain
\begin{eqnarray}
{\tilde E}_N
=
{\tilde E}_N^{(0)}
+
2
\omega
\sum_{\alpha=1}^r
\sum_{j=1}^{M_{\alpha}} I^{(\alpha)}_j,
\end{eqnarray}
where
${\tilde E}_N^{(0)}$
was given by the formula (\ref{nega-gra-ena}).
The quantity
$\sum_{\alpha=1}^r
\sum_{j=1}^{M_{\alpha}} I^{(\alpha)}_j$
which labels the excited states
takes values in the even non-negative integers.

Finally, we touch upon the operator
method.
We define operators
\begin{eqnarray}
  \label{a-c-operators}
{\cal B}_n
&=&
\frac{1}{2}
\sum_{\mu,\nu=1}^{2N}
\left(
({\bf L}^{-})^{n}
\right)_{\mu\nu}
=
\sum_{j,k=1}^N
\left(
({\cal A}^{--})^{\frac{n}{2}}
\right)_{jk},
\\
{\cal B}_n^{\dagger}
&=&
\frac{1}{2}
\sum_{\mu,\nu=1}^{2N}
\left(
({\bf L}^{+})^{n}
\right)_{\mu\nu}
=
\sum_{j,k=1}^N
\left(
({\cal A}^{++})^{\frac{n}{2}}
\right)_{jk},
\end{eqnarray}
where
$n\in \{2,4,6,\cdots\}$.
Note that
$\sum_{\mu,\nu}
\left(
({\bf L}^-)^n
\right)_{\mu,\nu}
=0$
for odd $n$.
For example,
$
{\cal B}_2^{\dagger}
=
{\cal H}_2
-
2\omega^2
\sum_{j=1}^N q_j^2
+
i\omega
\sum_{j=1}^N
(q_jp_j+p_jq_j)
$.
These satisfy the commutation relations,
\begin{eqnarray}
  \label{comm-a-c}
\left[
{\cal H}_2,
{\cal B}_n^{\dagger}
\right]
=
2n\omega{\cal B}_n^{\dagger},
\
\left[
{\cal H}_2,
{\cal B}_n
\right]
=
-2n\omega{\cal B}_n,
\end{eqnarray}
{\it i.e.},
these change the energy eigen value by $2n\omega$.
Let
$\left| 0
\right>
$ be the ground state of
the $B_N$-CSC model
with the ground state energy
${\bar E}_N^{(0)}$.
Then we can see that
the excited state
${\cal H}_2
\left|
{\bar \Psi}(\{n_j\})
\right>
=
{\bar E}_N
\left|
{\bar \Psi}(\{n_j\})
\right>
$
where ${\bar E}_N={\bar E}_N^{(0)}+
2\omega\sum_j n_j$
is constructed by
$
\left|
{\bar \Psi}(\{n_j\})
\right>
=
\prod_{j}
{\cal B}_{n_j}^{\dagger}
\left| 0
\right>
$.
Again, the number $\sum_j n_j$ is the even
non-negative integer.

We can conclude that the above three
methods are consistent
and give the same results.
The correlations via the
$1/r^2$-type interaction
appear only
in the ground state energy,
and excitations
with the fixed number of
electrons do not
include
any effects of
interactions.
Full details will be given in the
separate publication.

Finally, we comment on the Yangian symmetry
of our model.
We introduce the Dunkl operators\cite{dunkl}
\begin{equation}
  \label{dunkl}
D_j
=
\frac{\partial}{\partial q_j}
-
\lambda
\sum_{l=1, l\ne j}^N
\left(
\frac{1}{q_j-q_l}K_{jl}
+
\frac{1}{q_j+q_l}{\bar K}_{jl}
\right)
-
\lambda_1
\frac{1}{q_j}K_j,
\ (j=1, 2, \cdots, N).
\end{equation}
Here
$K_j, K_{jk}$ and
${\bar K}_{jk}$
are defined by
$
K_jq_j=-q_jK_j,\
K_{jk}q_k=q_jK_{jk},
$
and
${\bar K}_{jk}=K_jK_kK_{jk}$.
These satisfy the same relations among
$P_j, P_{jk}$ and
${\bar P}_{jk}$.
We find that the Dunkl operators satisfy
the relations,
$
K_{j}D_j
=
-D_jK_j,\
K_{jk}D_k
=
D_jK_{jk},\
{\bar K}_{jk}D_k
=
-
D_j{\bar K}_{jk}, and
$
\begin{eqnarray}
  \label{relation-dunkl}
\left[
D_j, D_k
\right]
&=&
0,
\\
\left[
q_j, D_k
\right]
&=&
\djk
\left(
1+
\lambda\sum_{l=1, l\ne j}^N
(
K_{jl}
+
{\bar K}_{jl}
)
+
2\lambda_1K_j
\right)
-
(1-\djk)\lambda
(K_{jk}
-
{\bar K}_{jk}).
\end{eqnarray}
Then,
following Refs. \cite{B-P-S,B-G-H-P,Hikami-a,B-H-W},
we can construct a monodromy matrix
which represents the Yangian symmetry of
the $B_N$-CSC model.
Details will be published elsewhere.

In this letter,
we have studied
the multicomponent Calogero model
of $B_N$-type
confined in the harmonic potential.
We found that
the model have conserved operators
as in other one-dimensional models with the
$1/r^2$-type interaction.
We have constructed the 'Jastrow'-type ground state of
our model.
Moreover,
we have obtained the exact spectrums.
For the model associated with
the Weyl group of type
$D_N$,
corresponding
quantities  are obtained from
the $B_N$ case
by setting
$\lambda_1=0
$.

The parallel constructions
hold for the
Sutherland model of
$BC_N$-type\footnote{%
The integrability of this model (with $\lambda_1'=0$)
has been studied by Etingof and Styrkas\cite{E-S}
in terms of the representation theory
of the Lie algebras. See also \cite{Chere2,van}.
}
(the spin Sutherland model of $BC_N$-type ($BC_N$-SS model)),
{\it i.e.}, the periodic model with
Hamiltonian\footnote{%
Using the identity
$\sin 2x=2\sin x\cos x$,
we rewrote the term $1/\sin^2(\pi/L)2q_j$.
};
\begin{eqnarray}
{\cal H}_3=\sum_{j=1}^N p_j^2
&+&
2
\bnor
\sum_{1\leq j<k \leq N}
\left\{
\frac{\lambda(\lambda-P_{jk})}{\displaystyle{
\sin^2\bnr(q_j-q_k)}}
+\frac{\lambda(\lambda-{\bar P}_{jk})}{\displaystyle{
\sin^2\bnr(q_j+q_k)}}
\right\} \nonun \\
&+&
\bnor
\sum_{j=1}^N
\left\{
\frac{\lambda_1(\lambda_1-P_j)}{\displaystyle{
\sin^2\bnr q_j}}
+
\frac{\lambda_1'(\lambda_1'-P_j)}{\displaystyle{
\cos^2\bnr q_j}}
\right\},
  \label{b-tri-hamiltonian}
\end{eqnarray}
where
$L$ is the linear size of the system.
For example,
${\cal H}_3$ is factorized as
\begin{equation}
  \label{suth-fact}
{\cal H}_3
=
\sum_{j=1}^N
b_j^{\dagger}b_j
+
{\cal F},
\end{equation}
where
\begin{eqnarray}
  \label{hhh}
b_j
&=&
-i\frac{\partial}{\partial q_j}
+
i\lambda\frac{\pi}{L}\sum_{l=1,l\ne j}^N
\left\{
\cot\frac{\pi}{L}(q_j-q_l)P_{jl}
+
\cot\frac{\pi}{L}(q_j+q_l){\bar P}_{jl}
\right\}
\nonumber
\\
& &
+
i\lambda_1\frac{\pi}{L}
\cot\frac{\pi}{L}q_jP_j
-
i\lambda_1'\frac{\pi}{L}
\tan\frac{\pi}{L}q_jP_j ,
\\
{\cal F}
&=&
\left(\frac{2\pi}{L}\right)^2
\left[
\frac{\lambda^2}{2}N(N-1)
+
\frac{\lambda^2}{12}
\sum_{j,k,l=1,j,k,l:\mbox{{\tiny distinct}}}^N
(
P_{jk}P_{kl}
+
{\bar P}_{jk}{\bar P}_{kl}
+
P_{jk}{\bar P}_{kl}
+
{\bar P}_{jk}P_{kl}
)
\right.
\nonumber
\\
& &
\left.
+
\frac{(\lambda_1+\lambda_1')^2}{4}N
+
\frac{\lambda(\lambda_1+\lambda_1')}{4}
\sum_{j,k=1,j\ne k}^N
(P_j+P_k)P_{jk}
\right].
\end{eqnarray}
Also,
the
discritization \cite{S-A94,B-P-S,lat-conf1,lat-conf2}
of the $B_N$-CSC model
and the $BC_N$-SS model can be constructed.

The models
which were introduced in this letter
may become useful for
the one-dimensional physics
in the presence of boundaries,
in particular, with internal degrees of freedom.
For example,
those models will be helpful in explaining
the internal structure of the FQHE
state\cite{B-W,Read}.
We hope to turn to this issue
in the near future.

{\bf Acknowledgement}\\
I wish to thank N. Kawakami for critical reading
of the manuscript.
This work was supported by the Yukawa memorial foundation.


\end{document}